\documentclass[conference]{IEEEtran} 
\usepackage{amsmath,amssymb}
\usepackage{mathtools}
\usepackage{graphicx,epsfig}
\usepackage{svg}
\usepackage{multirow}
\usepackage{mwe}

\begin{document}

\title{Low Latency Decoder for Short Blocklength Polar Codes}
\author{ \IEEEauthorblockN{{Heshani~Gamage,~Vismika~Ranasinghe,~Nandana~Rajatheva,~and~Matti~Latva-aho.~}}
\IEEEauthorblockA{Centre for Wireless Communications,~~University of Oulu, Finland\\
E-mail: \{ heshani.niyagamagamage, vismika.maduka, nandana.rajatheva, matti.latva-aho \} @oulu.fi}
}
\maketitle
\vspace{-0.2pt}

\begin{abstract}
Polar codes have been gaining a lot of interest due to it being the first coding scheme to provably achieve the symmetric capacity of a binary memoryless channel with an explicit construction. However, the main drawback of polar codes is the low throughput of its successive cancellation (SC) decoding. Simplified SC decoding algorithms of polar codes can be used to reduce the latency of the polar decode by faster processing of specific sub-codes in the polar code. By combining simplified SC with a list decoding technique, such as SC list (SCL) decoding, polar codes can cater to the two conflicting requirements of high reliability and low latency in ultra-reliable low-latency (URLLC) communication systems. Simplified SC algorithm recognises some special nodes in SC decoding tree, corresponding to the specific subcodes in the polar code construction,  and efficiently prunes the SC decoding tree, without traversing the sub-trees and computing log-likelihood ratios (LLRs) for each child node. However, this decoding process still suffers from the latency associated with the serial nature of SC decoding. We propose some new algorithms to process new types of node patterns that appear within multiple levels of pruned sub-trees and it enables to process certain nodes in parallel.In short blocklength polar codes, our proposed algorithm can achieve up to $13\%$ latency reduction from fast-simplified SC \cite{Sarkis2013} without any performance degradation. Furthermore it can achieve up to $27\%$ latency reduction if small error-correcting performance degradation is allowed.  
\end{abstract}


\section{Introduction}
Introduced by Arikan in \cite{Arikan2009a}, polar codes are the first channel coding scheme to provably achieve the symmetric capacity of a binary memoryless channel with low encoding and decoding complexities. In addition, they remain the only capacity achieving block code with an explicit construction. However, a primary concern in regard to polar codes is the long decoding latency of its SC decoding algorithm due to the serial nature of processing. Since the introduction of polar codes and the SC decoding algorithm, several algorithms have been proposed to improve the decoding performance. 

SC-list (SCL) \cite{Tal2015}, SC-stack (SCS) \cite{Niu2012a}, and SC-flip (SCF) \cite{Afisiadis2015} decoding algorithms are based on the concept of list search. Given enough list/stack or flip size, they can achieve near maximum-likelihood (ML) performance at a cost of high computational and memory complexities. In addition to one of SCL, SCS or SCF methods, a cyclic redundancy check (CRC) can be added to the code to aid the selection of the most likely codeword from the list\cite{Niu2012b}, which helps in improving the bit error performance even above ML performance. CRC can be added to the code without affecting the code-rate by utilizing frozen bit positions of the polar code to encode the CRC bits. Since all of these methods based on SC algorithm are processed serially, there are limitations in speeding up the decoding process, restricting the decoder throughput. In \cite{Li2013} parallel decoders are implemented by splitting the generator matrix into component codes and processing them in parallel with small performance degradation.  Distributing CRC bits \cite{Chen2017} and  partitioning the codeblock \cite{Hashemi2018} are proposed to speed up the decoding process in SCL decoding.

On the other hand, several investigations have been carried out to simplify the SC algorithm to make it faster, without reducing the error-correcting performance. These techniques are generally based on identifying prevalent sub-trees in SC decoding binary tree and efficiently pruning these sub-trees for faster decoding. In simplified SC (SSC) \cite{Alamdar-Yazdi2011} algorithm, rate-0 (R0) and rate-1(R1) nodes are introduced to prune the binary decoding tree. In \cite{Sarkis2013} ML-SSC algorithm, ML nodes are introduced to further prune the binary decoding tree to speed up the decoding process. In \cite{Sarkis2014} fast-SSC algorithm, two more types of nodes are introduced, namely single parity check (SPC) and repetition (REP) nodes for further pruning the binary tree. In \cite{Hanif2017}, some of more specific bit patterns are identified as new nodes for the decoder tree-pruning. Occurrence of some of these nodes is not very common in short blocklength codes. Generalized-fast algorithm \cite{Condo2019} generalizes some of the nodes categorized in \cite{Hanif2017} to broader categories and provides algorithms of efficient mergers for some of the special nodes mentioned above.

In this paper, we propose a faster decoding algorithm for increasing the throughput of SC decoding based on the most prominent node patterns that appear in short blocklength polar codes, within multiple levels of the binary decoding tree. We introduce an algorithm to parallelize the processing of some node patterns to avoid the latency associated with sequential nature of SC decoding. We compare the error performance and the complexity of the proposed algorithm with the fast-SSC algorithm.

The rest of this paper is organized as follows. In section II, we review some preliminary material of polar codes, SC decoding algorithm and other fast SC decoding algorithms. In section III, we propose a new algorithm for improving the throughput of SC decoder . In section IV, we compare the bit error rate (BER), frame errror rate (FER) performances and the latency of the proposed algorithm with the fast-SSC algorithm. Finally in section V we provide some concluding remarks.


\section{Preliminaries}
\subsection{Polar Codes}
A binary polar code of length $N=2^n$ with $K$ information bits and rate $R \triangleq K/N$ is denoted by $\mathcal{P}(N,K)$. It can be constructed by concatenating two polar codes of length $N/2$. The recursive construction process can be represented by modulo-2 matrix multiplication as
\renewcommand{\vec}[1]{\mathbf{#1}}
\begin{equation}
    \vec{x} = \vec{u} \vec{G}^{\otimes n}
\end{equation}
where, $\vec{u} = \{ u_0, u_1, ... u_{N-1} \}$ is the sequence of input bits containing information bits and frozen bits, $\vec{x} = \{ x_0, x_1, ... x_{N-1} \}$ is the encoded vector, and the code generator matrix $\vec{G}^{\otimes n}$ is the $n$-th Kronecker product of the polarizing matrix $\vec{G} = \begin{bsmallmatrix}1&0\\1&1\end{bsmallmatrix}$. 

The concept of channel polarization attributed to polar codes is, transforming $N$ copies of a channel with a symmetric capacity of $I(W)$, into extreme channels of capacity close to one (completely reliable) or zero (completely noisy). Channel polarization is achieved through recursively applying a polarization transform $\vec{G}$. Out of $N$ channels, $I(W)$ fraction of channels will become perfectly reliable channels and $1-I(W)$ fraction of channels will become completely noisy channels. Then, the information bits are sent only through best $K$ synthetic channels while inputs to other $(N-K)$ channels are made ``frozen'' bits by setting it at a predefined value; one or zero, known at the decoder. Throughout this paper, we assume all the frozen bits are set at zero.  $\mathcal{A}$ denotes the set of information bit indices and $\mathcal{A}^c$ denotes the set of frozen bit positions which are known to the decoder.

\subsection{Successive Cancellation Decoding}
\begin{figure}
    \centering
    \includegraphics[width=1\columnwidth]{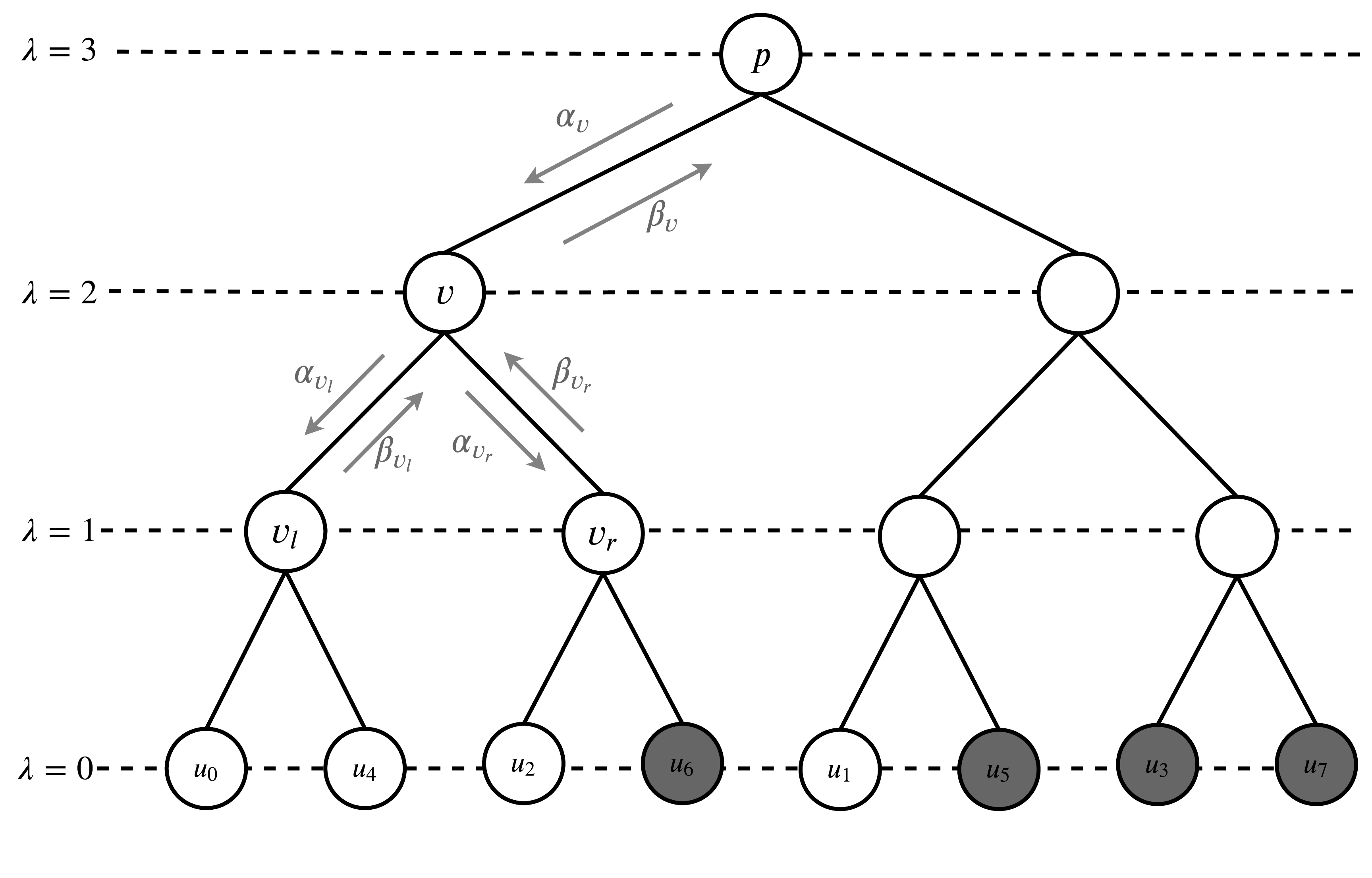}
    \caption{SC decoding tree structure for a polar code of block length $N = 8$ and rate $R = \frac{1}{2}$.}
    \label{fig:SCtree}
\end{figure}

SC based decoding algorithms can be represented through a binary tree. In Fig. \ref{fig:SCtree} the SC decoding binary-tree structure for a  polar code of length $N=8$ and rate $1/2$ is presented. Here, darkened leaves of the binary tree represent the information bit nodes and white leaves represent the frozen bits. The binary tree has $n+1$ levels where, $n = \log_2N$. $\lambda \in [0,n]$ denotes the level of the current processing node $v$ of the tree. At a given level $\lambda$ there are $2^{n-\lambda}$ nodes, and each node at level $\lambda$ has $2^\lambda$ leaves. A vector of size $2^{\lambda}$ is exchanged through the branch between nodes at levels $\lambda$ and $\lambda+1$ during the traversal of the binary-tree. It should be noted that the decoding is done in bit-reversed order as in the butterfly diagram of \cite{Arikan2009a}. Here, the right leaf of a node is $2^{n-1}$ indices away from the index of the corresponding left leaf.

Input to the decoder is an LLR vector $\Vec{\alpha_n = \{ \alpha _0, \alpha _1, ... \alpha _{N-1}\}}$ at the root node (at level $n$). At each level $\lambda$, a $2^\lambda$ length LLR vector $\alpha$ is coming to the node as input.

Once the node at level $\lambda$ receives the LLR vector $\alpha _v$, it calculates the $2^{(\lambda-1)}$ length LLR vector $\alpha_{v_l}$ as
\begin{equation}\label{eq:left_process}
    \alpha_{v_l}[i]=\alpha_v[2i] \boxplus \alpha_v[2i+1] ~ \textrm{for} ~i \in [0:2^{(\lambda-1)}-1].
\end{equation}
 and passes to the left child node $v_l$. Here, the binary operator $\boxplus$ denotes the operation
 \begin{equation}\label{eq:boxplus}
     x \boxplus y = 2 atanh (\tanh \frac{x}{2} \tanh \frac{y}{2} ).
 \end{equation}
In order to reduce the complexity, Eq.\ref{eq:boxplus} can be approximated by min-sum simplification \cite{Leroux:2012}. 
The local decoder node $v$ then waits until it receives hard bits vector $\beta _{v_l}$ from the left child node $v_l$, and calculates $\alpha_{v_r}$ as
\begin{equation}\label{eq:right_process}
    \alpha_{v_r}[i]=\alpha_v[2i] (1-2\beta_{v_l}[i]) + \alpha_v[2i+1] ~ \textrm{for} ~i \in [0:2^{(\lambda-1)}-1].
\end{equation}
and passes to the right child node $v_r$. Once it receives the hard bits vector $\beta _r$ from the right child node, it calculates the codeword $\beta_v$ as
\begin{equation}
    \beta_{v}[2i]=\beta_{v_r}[i] \oplus \beta_{v_l}[i],
\end{equation}
\begin{equation}
    \beta_{v}[2i+1]=\beta_{v_r}[i].
\end{equation}
for $i \in [0:2^{(\lambda-1)}-1]$. If the $v$ current processing node is a leaf node at level $0$, once it receives the $\alpha _v$, it calculates the $\beta _v$ as
\begin{equation}
    \beta _v =     \begin{cases}
      h(\alpha _v); & \text{if}\ v \in \mathcal{A} , \\
      0; & \text{if}\ v \in \mathcal{A}^c .
    \end{cases}
\end{equation}
$h(x)$ is the binary quantizer with
\begin{equation}
    h(x) =     \begin{cases}
      0  ; & \text{if}\ x \geq 0, \\
      1  ; & \text{otherwise.}
    \end{cases}
\end{equation}

\subsection{Fast SC Decoding}
In order to increase the speed of SC based decoding, particular sequences of frozen and information bit patterns have been identified from the leaves of the binary tree. Efficient fast decoders have been proposed based on performing tree-pruning on these nodes. We denote a frozen bit as `0' and information bit as `1' in the polar code construction pattern $\vec{s}$. The pattern for the polar code in Fig.\ref{fig:SCtree} can be written as $\vec{s} = \{0, 0, 0, 1, 0, 1, 1, 1\}$. 

According to the polar construction, bits which are decoded first tend to be of lower reliability than bits that are decoded later in the decoding tree. Therefore, the frozen bits tend to concentrate into first $K/N$ leaves of the binary tree and information bits tend to be concentrated in the end of the tree with a grey area between concentrated frozen and information bits.  

\subsubsection{Simplified Nodes}
The following nodes which are introduced in \cite{Alamdar-Yazdi2011} and \cite{Sarkis2013} are the most frequent simplified nodes occur in short block length polar code construction patterns. 
\begin{itemize}
  \item \textit{R0 node}:  A node at level $\lambda$  where all the $2^\lambda$ corresponding leaf nodes of the sub-tree are frozen which can be denoted by $ \vec{s} = \{0, 0, ... , 0\}$. Then the tree can be pruned at the rate-0 node and code vector can be set as $\beta_v [i]= 0 ~\text{for}~ i \in [0, 2^\lambda-1 ]$ 
  \item \textit{R1 node}: A node at level $\lambda$, where all the $2^\lambda$ corresponding leaf nodes of the sub-tree are information bits. This can be denoted by $ \vec{s} = \{1, 1, ... , 1\}$. Then the tree can be pruned at the R1 node and code vector can be set as $\beta_v [i]= h(\alpha_v[i]) ~\text{for}~  i \in [0, 2^\lambda-1 ]$.
  \item \textit{REP node}: A node at level $\lambda$  with the last right leaf is an information bit and  all the other  $2^\lambda-1$ leaf nodes of the sub-tree are frozen,  which can be denoted by the pattern $ \vec{s} = \{0, 0, ..., 0, 1\}$ . Then the tree can be pruned at the REP node and code vector can be set as $\beta_v [i]= h( \sum_{i=0}^{2^\lambda-1} \alpha_v[i]) ~\text{for}~ i \in [0, 2^\lambda-1 ]$.
  \item \textit{SPC node}:  A node at level $\lambda$  with the first left leaf is a frozen bit and all the other  $2^\lambda-1$ leaf nodes of the sub-tree are information bits, which can be denoted by the pattern $ \vec{s} = \{0, 1, ...,1\}$. 
 Here, hard decisions of the of the LLR vector is calculated as $h(\vec{\alpha_v})$ and the parity bit is calculated as 
 \begin{equation}
     \text{parity} = \sideset{}{h(\alpha_v[i])}\bigoplus_{i=0}^{ 2^\lambda-1}
     \label{eq_spc_parity}
 \end{equation} Then the index of the least reliable bit is found from 
 \begin{equation}
     j = \underset{i}{\arg \min }~|\alpha_v[i]|.
     \label{eq_spc_min}
 \end{equation}
 Then the output of the node can be calculated as 
 \begin{equation}
     \beta_v [i]= \begin{cases}
 h( \vec{\alpha_v}) \oplus \text{parity} &\text{when $i=j$}\\
 h( \vec{\alpha_v})  &\text{otherwise}
\end{cases}
\label{eq_spc_flip}
 \end{equation} 
for  $ i \in [0, 2^\lambda-1 ]$. 
\end{itemize}

\subsubsection{Simplified Node Mergers}
Theses are the mergers between simplified nodes mentioned above for further reducing the latency.  
\begin{itemize}
    \item \textit{REP-SPC Merge} \cite{Sarkis2014}:
    This merge is achieved by having two SPC decoders, $SPC_0$ and $SPC_1$, whose inputs are calculated assuming output of the REP code is 0 and 1 respectively. 
    \item \textit{Generalized REP (G-REP) Merge} \cite{Condo2019}:
    This is an extension of node mergers in \cite{Sarkis2014}, where the nodes in multiple levels are merged. This is a node at any stage $L$ whose descendants are \textit{R0} nodes except the rightmost node at a certain stage $l_0 < L$, which is a generic node of rate $C$. 
    \item \textit{Generalized Parity-Check (G-PC) Merge} \cite{Condo2019}: 
    This is a generalized version of SPC nodes of \cite{Sarkis2014}. This is a node at stage $L$ in which all its descendants are $R1$ nodes except the leftmost node at level $l_0 < L$, which is an $R0$ node.
\end{itemize}
\smallskip 
After pruning the tree according to the simplified nodes as mentioned above, hard-bits are propagated until the root node unlike in the SC algorithm. After we get the $\beta_0$ at the root node, we can calculate the $\hat{x}$ from
\begin{equation}
    \hat{x} = \beta_0 \vec{G}.
\end{equation}
We can get the bit estimates $\hat{u}$ from $\hat{x}$ using the construction pattern $\vec{s}$.

\section{Proposed Algorithm} 
\begin{figure*}[t]\centering
    \centering
    \includegraphics[width=2\columnwidth]{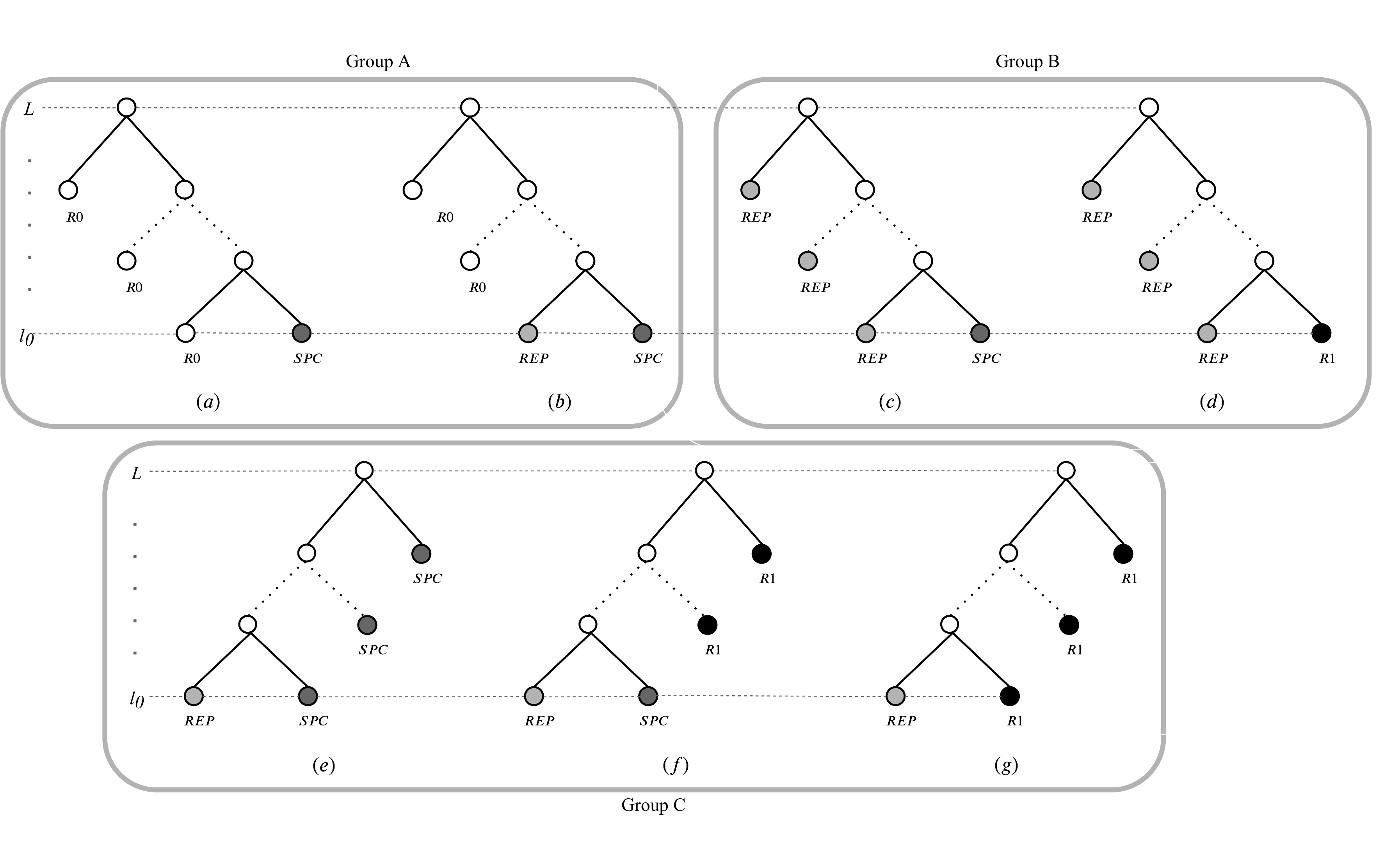}
    \caption{Most frequent node patterns in short blocklength polar codes.}
    \label{fig:node_patterns}
\end{figure*}

 We analysed the most frequent nodes and node patterns in short polar codes and we observed that only a limited number of patterns are prevalent. Node patterns denoted in the binary sub-trees $(a)$ to $(g)$ in Fig. \ref{fig:node_patterns} are the most prominent patterns in short blocklength polar codes. We propose  efficient algorithms for processing these node patterns in multiple levels of the binary-tree. Our algorithm enables processing of the several left-most nodes of the binary sub-tree in parallel at the node merging point, so that it will avoid the serial nature of the fast-SSC.
 
Here, we assume the root node of the sub-tree is at level $L$ and the leaf nodes (pruned) are at level $l_0$. Therefore, the depth of the sub-tree is given by, $t = L-l_0$. We can group the node patterns (a) to (g) in Fig. \ref{fig:node_patterns} into three groups as bellow. 

\subsection{Group A patterns}
Patterns $(a)$ and $(b)$ of Fig. \ref{fig:node_patterns} falls under $G-REP$ merge in \cite{Condo2019}. In the pattern (a), There are $t$ $R0$ nodes from level $L$ to leaf level $l_0$ and an $SPC$ node as the rightmost leaf. We name this as $R0^t-SPC$ and In the pattern (b), There are $t-1$ $R0$ nodes from level $L$ to level $l_0 + 1$, and a $REP-SPC$ node as rightmost sub-tree at level $l_0+1$. We name this as $R0^{t-1}-REP-SPC$ pattern. 
Pattern $(a)$ can be identified as a $G-REP$ node with $Rate-C$ node is replaced with $REP$ node. In the pattern $(b)$,  $Rate-C$ node of $G-REP$ node is replaced with a $REP-SPC$ merged node. 

\subsection{Group B patterns}
In patterns $(c)$ and $(d)$ of Fig.\ref{fig:node_patterns} , there are $t$ $REP$ nodes from level $L-1$ to leaf level $l_0$. In pattern $(c)$, the right-most leaf node is an $SPC$ node whereas in pattern $(d)$ the rightmost leaf is an $R1$ node. We name these as $REP^t-SPC$ and $REP^t-R1$ respectively.  These are the most frequently appearing node patterns in the most polar code construction patterns. 

$REP^t-SPC$ node merger can be processed faster as follows. Assuming the information bit at a REP node at level $l$ is $q_l$, We first calculate the information bit at each REP node at level $l$ in parallel as
\begin{equation}
    q_l = h(\sum_{k=0}^{2^{l}-1} \sum_{k=0}^{2^{t}-1}  \alpha_{2^{t}i+k} \boxplus  \alpha_{2^{t}i+k+2^{t-1}}).
\end{equation}

After calculating $t$ $REP$ nodes in parallel, the decoded information bits are encoded again before decoding the $SPC$ node. First information bits to encode are transformed as a concatenation of $t$ $REP$ nodes of size $2^{t-1}$ to 1 (size one REP node is equal to $q_L$). The nodes are in the order from the lowest level node to the highest level node. Last bit is set as $0$. For example for $t=3$, the sequence to be encoded is
\begin{equation}
     q = \{0, 0, 0, q_{l_0}, 0, q_{l_0+1}, q_{l_0+2}, 0\}.
\end{equation}
This is encoded using a polar code generator matrix of size $2^{t-1}$ as
\begin{equation}
    a = qG.
\end{equation}
Now the encoded $SPC$ bits from the $SPC$ node can be directly calculated as 
\begin{equation}
 \beta_i = h(\sum_{k=0}^{2^{t}-1} (1-2a_k)\alpha_{2^{t}i+k}) ~\text{for}~i \in ~\{0, 2^{l_0}-1\}.
\end{equation}
For the final encoded partial sum bits going out from the merge node at level $L$, encoded $a$ bits are added to each $\beta_i$ as $a \oplus \beta_i$ for $i$ $\in$ $\{0, 2^{l_0}-1\}$

$REP^t-R1$ node merge also can be processed in exactly the same procedure as above $REP^t-SPC$ node, replacing the $SPC$ node with an $R1$ node. Both of these node mergers will case a small degradation in error-correcting performance of the code. 

\subsection{Group C patterns}
In patterns $(e)$ , $(f)$and $(g)$ of Fig.\ref{fig:node_patterns} , There is a $REP$ node at the leftmost leaf of the pruned binary subtree. In the pattern (e), all the rightmost children are $SPC$ nodes from level $l_0$ to level $L-1$. In pattern $(f)$, all the right-most child nodes are $R1$ nodes except the right-most leaf node at the level $l_0$ which is an $SPC$ node. In pattern $(g)$  all the right child nodes are $R1$ nodes. We name these patterns as $REP-SPC^t$, $REP-SPC-R1^{t-1}$, and $REP-R1^t$ respectively.  
The $REP-SPC^t $ merge can be made faster by the following algorithm, at the expense of a small performance loss. 
First the $REP$ node is LLR is decoded as
\begin{equation}
    q_L = h(\sum_{i=0} ^{2^{l_0}-1} \sum_{j=0}^{2^t-1} \boxplus ~{\alpha _{2^ti+j}}).
\end{equation}
Now partial sum bits at level $L$ can be directly calculated in parallel from
\begin{equation}
    \begin{split}
        \beta_{2^ti+k} = h(\alpha_{2^ti+k} + \sum_{j=0/k}^{2^t-1} \boxplus ~ \alpha_{2^ti+j}), \\
        \text{~for~} i \in ~\{0, 2^{l_0}-1\}  \text{~and~}  k \in  ~\{0, 2^{t}-1\}. 
    \end{split}
\end{equation}

Finally $2^t$ parity checks can be performed for each $i$ such that
\begin{equation}
     \sum_{k=0}^{2^{l_0}-1}\beta_{2^ti+k} + q_L = 0.
\end{equation}
When the parity check is not satisfied, the partial sum bit with the least reliable LLR value can be flipped similar to the $SPC$ node processing.

The $REP-SPC-R1^{t-1} $ node merger can also be made faster following the procedure above and $REP-R1^t $ node can be decoded similar to $REP-SPC^t $, without the final parity check.

\section{Performance} 
In this section we first compare the decoding latency  between  the fast-SSC algorithm \cite{Sarkis2014} and the proposed multi-level mergers. Similar to the work in \cite{Condo2019}, we assume equations (\ref{eq:left_process}), (\ref{eq:right_process}), and  $R0$, $R1$ nodes have a cost of 1 time step each. Furthermore, processing of  $REP$ and $SPC$ nodes have costs of 2 and 3 time steps respectively. TABLE: \ref{tab:merge_latency} tabulates the latency for each node merger in terms of time steps. Since our focus is only on reducing the decoding latency, we assume unlimited resource availability for the latency calculations. This enables the decoder to process in parallel whenever possible to achieve minimum latency.

\begin{table}[h]
\centering
\begin{tabular}{|c|c|c|}
\hline
Merge & Time steps\\ \hline
$R0^t-SPC$        & 4         \\ \hline
$R0^{t-1}-REP-SPC$ & 4 \\ \hline
$REP^t -SPC$       & 9          \\ \hline
$REP^t -R1$        & 8         \\ \hline
$REP-SPC^t$        & 7          \\ \hline
$REP-SPC-R1^{t-1}$    & 7           \\ \hline
$REP-R1^t$      & 7          \\ \hline
\end{tabular}
\caption{Number of time steps assumed for different node mergers.}
\label{tab:merge_latency}
\end{table}

In TABLE: \ref{table:complexity_with_nodes}, obtained latency improvements for polar codes of blocklengths $128$ and $512$ and rates of $1/2$ and $1/4$ are presented. Here, we calculate the latency in terms of time steps for fast-SSC, proposed algorithms employing only the loseless mergers, and proposed algorithms for all the mergers. It can be seen that we can achieve up to $13\%$ of latency reduction compared to fast-SSC by using only the lossless mergers of proposed algorithm . Furthermore, we can achieve up to $27\%$ of latency reduction using all the proposed mergers at a cost of a small degradation in error performance.   

\begin{figure*}[t]\centering
    \centering
    \includegraphics[width=2\columnwidth]{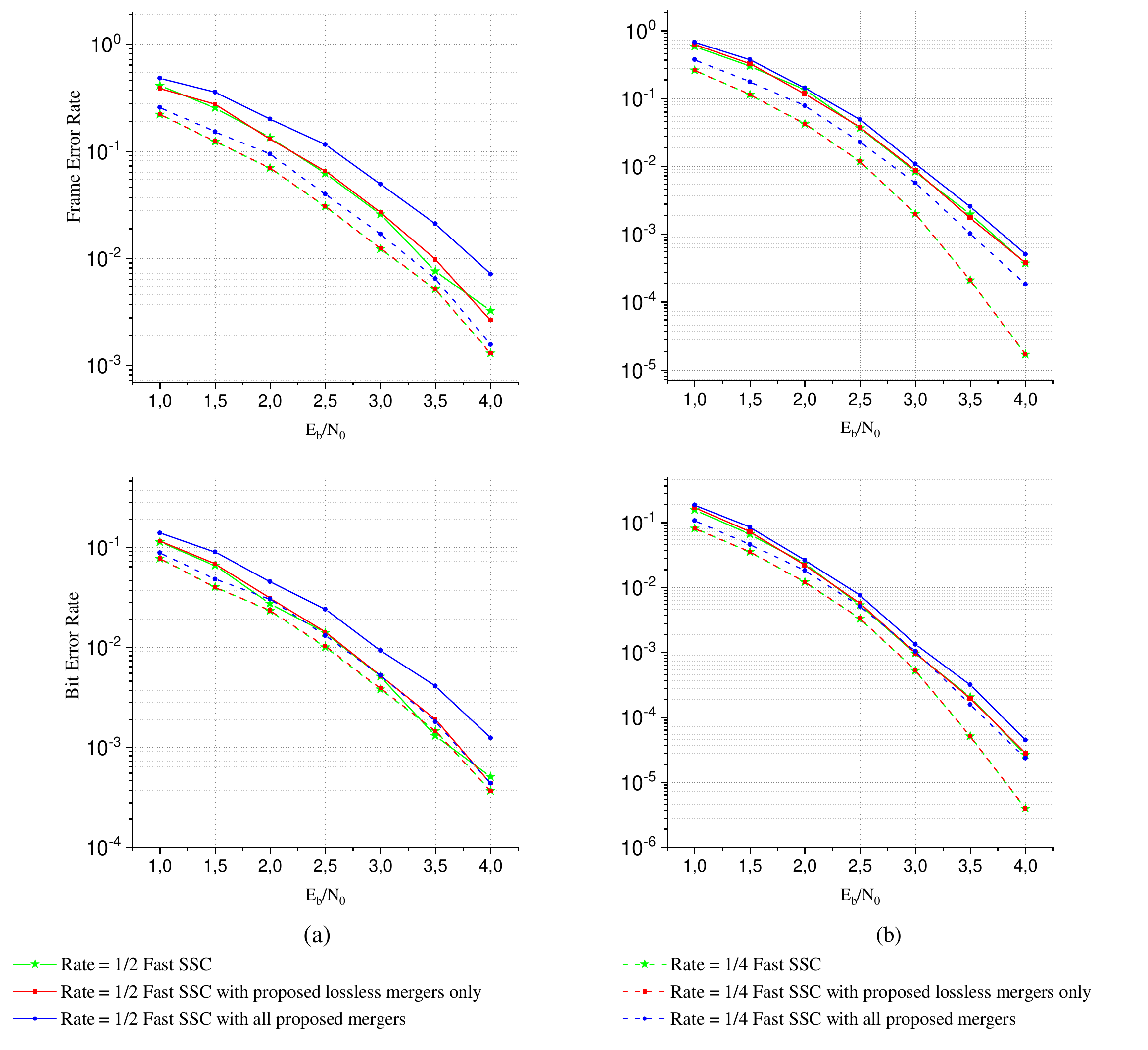}
    \caption{FER and BER performance of the proposed algorithm for polar codes of length $(a) N=128$ $(b) N=512$ and for rates $R=1/4$ and $1/2$,  with and without lossy mergers.}
    \label{fig:results}
\end{figure*}

Fig. \ref{fig:results} compares the error correcting performance of the  fast-SSC decoder and proposed mergers for polar codes of block lengths $N = 128$ and $512$ at $R = 1/2$ and $1/4$. It should be noted that the BER and FER performances of fast-SSC decoder are similar to that of the SC decoder. It can be observed that for $R0^t-SP$C, $REP-R1^{t-1}$, and $R0^t-REP-SPC$ mergers the error correcting performance is similar to that of fast-SSC. Hence, for those nodes, as shown in TABLE:  \ref{table:complexity_with_nodes}, further improvements in latency can be achieved without sacrificing the error correcting performance.  

However, a degradation of error performance can be observed for mergers $REP^t-R1$, $REP^t -SPC$, $REP-SPC^t$, and $REP-SPC-R1^{t-1}$. In $REP^t -SPC$ nodes, the performance degradation is caused by calculating the REP nodes in parallel as it alters the optimality of successive cancellation decoding. As the number of $REP$ nodes in the merger increases, the impact on error correcting performance is greater. For the $REP-SPC^t$ and $REP-SPC-R1^{t-1}$ mergers, since a single parity check is used after calculating the REP bit, performance degradation is caused by ignoring the constraints imposed by $SPC$ nodes in merger. This is similar to the idea of ignoring the frozen bits in $Rate-C$ node in $G-REP$ merger to achieve a better latency in \cite{Condo2019}. These ignored frozen bits are known as addition frozen bits(AF) bits. Hence larger the number of SPC nodes in the merger, greater impact on performance.

\begin{table}[h]
\renewcommand{\arraystretch}{1.35}
\resizebox{\columnwidth}{!}{
\begin{tabular}{|c|c|c|c|c|}
\hline
Parameters                                                                     & \begin{tabular}[c]{@{}c@{}}Simulation\\ name\end{tabular} & Enabled mergers                                                                                                                                                                                                                                                                                   & Complexity  &\begin{tabular}[c]{@{}c@{}}Latency\\ reduction\end{tabular} \\ \hline
\multirow{3}{*}{\begin{tabular}[c]{@{}c@{}}$N = 128$\\ $R = \frac{1}{2}$\end{tabular}}     & Fast-SSC                                                       & -                                                                                                                                                                                                                                                                                                 & 55  &  -       \\ \cline{2-5} 
                                                                               & Proposed loseless mergers                                                  & \begin{tabular}[c]{@{}c@{}}$R0^t-SPC$\\ $REP-R1^t$\end{tabular}                                                                                                                                                                                         & 49  & 11\%       \\ \cline{2-5} 
                                                                               & All proposed mergers                                                     & \begin{tabular}[c]{@{}c@{}}$R0^t-SPC$\\ 
                                                                               $REP-SPC^t$\\ 
                                                                               $REP^t -SPC$\\ 
                                                                               $REP-R1^t$\end{tabular}                                                                                                                   & 42   &   24\%   \\ \hline
\multirow{3}{*}{\begin{tabular}[c]{@{}c@{}}$N = 128$\\ $R = \frac{1}{4}$ \end{tabular}} & Fast-SSC                                                       & -                                                                                                                                                                                                                                                                                                 & 50     & -    \\ \cline{2-5} 
                                                                               & Proposed loseless mergers                                                  & -                                                                                                                                                                                                                                                                  & 50  & 0\%        \\ \cline{2-5} 
                                                                               & All proposed mergers                                                     & \begin{tabular}[c]{@{}c@{}}$REP^t -SPC$\\ $REP^t -R1$\end{tabular}                                                                                                                                                                                     & 41    & 18\%     \\ \hline
\multirow{3}{*}{\begin{tabular}[c]{@{}c@{}}$N = 512$\\ $R = \frac{1}{2}$\end{tabular}}     & Fast-SSC                                                       & -                                                                                                                                                                                                                                                                                                 & 167      & -  \\ \cline{2-5} 
                                                                               & Proposed loseless mergers                                                  & \begin{tabular}[c]{@{}c@{}}
                                     
                                                                               $REP-R1^t$\\ 
                                                                               $R0^t-SPC$\\ $R0^{t-1}-REP-SPC$\end{tabular}                                                                                                             & 145   & 13\%     \\ \cline{2-5} 
                                                                               & All proposed mergers                                                     & \begin{tabular}[c]{@{}c@{}}$REP^t -R1$\\ $REP-R1^t$\\ $R0^t-SPC$\\ $R0^{t-1}-REP-SPC$\\ $REP-SPC^t$\\ $REP^t -SPC$\\ $REP-SPC-R1^{t-1}$\end{tabular} & 130    & 22\%    \\ \hline
\multirow{3}{*}{\begin{tabular}[c]{@{}c@{}}$N = 512$\\ $R = \frac{1}{4}$ \end{tabular}}    & Fast-SSC                                                       & -                                                                                                                                                                                                                                                                                                 & 165   & -     \\ \cline{2-5} 
                                                                               & Proposed loseless mergers                                                  & \begin{tabular}[c]{@{}c@{}}$R0^t-REP-SPC$\\ 
                                                                               $R0^t-SPC$\end{tabular}                                                                                                                                                 & 145 &     12\%   \\ \cline{2-5} 
                                                                               & All proposed mergers                                                     & \begin{tabular}[c]{@{}c@{}}$R0^{t-1}-REP-SPC$\\ $REP^t -SPC$\\ $REP-SPC^t$\\ 
                                                                               $REP^t -R1$\\ $R0^t-SPC$\end{tabular}                                  & 120 & 27\%       \\ \hline
\end{tabular}
}
\caption{Latency improvements with proposed mergers for $N =[128, 512]$ and $R = [\frac{1}{4}, \frac{1}{2}]$}
\label{table:complexity_with_nodes}
\end{table}

\section{Conclusion}
In this work, we introduced new multi-level node mergers for fast decoding of short blocklength polar codes base on most frequent node patterns in polar code construction patterns. The proposed algorithm is evaluated for the latency reduction in terms of number of time steps for short blocklength polar codes. In addition, the error correcting performance of the proposed algorithm is compared with the fast-SSC algorithm. Our algorithm can gain up to $13\%$ of latency reduction without any performance degradation, using only proposed losless mergers and it is possible to achieve up to $27\%$ latency reduction if lossy mergers are allowed. This latency improvement is achieved through parallel processing of nodes in the mergers, avoding the serial nature of SC decoding. 

\bibliographystyle{IEEEtran} 
\bibliography{Sec_7_Bibliography.bib}  

\end{document}